\documentclass[11pt]{article}
\usepackage[dvips]{graphicx}
\usepackage{amssymb}
\usepackage{color}
\usepackage{amsmath}
\usepackage{nicefrac}
\usepackage[left]{lineno}
\usepackage{hyperref}
\usepackage{adjustbox}
\usepackage{xspace}
\usepackage{multirow}
\usepackage{caption}
\usepackage{subcaption}
\usepackage{graphicx}

\definecolor{burgundy}{rgb}{0.5, 0.0, 0.13}

\def\geant {\mbox{\textsc{Geant4}}\xspace}

\begin{document}
\centerline{\LARGE EUROPEAN ORGANIZATION FOR NUCLEAR RESEARCH}
%
\vspace{10mm} {\flushright{
CERN-EP-2021-018 \\
28 January 2021\\
\vspace{4mm}
Revised version:\\11 March 2021\\
}}
\vspace{-30mm}

%
%

%
\vspace{40mm}

\begin{center}
\boldmath
{\bf {\Large\boldmath{Search for $K^+$ decays to a muon and invisible particles}}}
\unboldmath
\end{center}
\vspace{1mm}
\begin{center}
{\Large The NA62 Collaboration}\\
\end{center}

\begin{abstract}
The NA62 experiment at CERN reports searches for $K^+\to\mu^+N$ and $K^+\to\mu^+\nu X$ decays, where $N$ and $X$ are massive invisible particles, using the 2016--2018 data set. The $N$ particle is assumed to be a heavy neutral lepton, and the results are expressed as upper limits of ${\cal O}(10^{-8})$ of the neutrino mixing parameter $|U_{\mu4}|^2$ for $N$ masses in the range 200--384 MeV/$c^2$ and lifetime exceeding 50 ns. The $X$ particle is considered a scalar or vector hidden sector mediator decaying to an invisible final state, and upper limits of the decay branching fraction for $X$ masses in the range 10--370 MeV/$c^2$ are reported for the first time, ranging from ${\cal O}(10^{-5})$ to ${\cal O}(10^{-7})$. An improved upper limit of $1.0\times 10^{-6}$ is established at 90\% CL on the $K^+\to\mu^+\nu\nu\bar\nu$ branching fraction.
\end{abstract}

\begin{center}
{\it Published as Physics Letters B816 (2021) 136259}
\end{center}

\newpage
\begin{center}
{\Large The NA62 Collaboration$\,$\renewcommand{\thefootnote}{\fnsymbol{footnote}}%
\footnotemark[1]\renewcommand{\thefootnote}{\arabic{footnote}}}\\
\end{center}
\vspace{3mm}
\begin{raggedright}
\noindent
{\bf Universit\'e Catholique de Louvain, Louvain-La-Neuve, Belgium}\\
 E.~Cortina Gil,
 A.~Kleimenova,
 E.~Minucci$\,$\footnotemark[1]$^,\,$\footnotemark[2],
 S.~Padolski$\,$\footnotemark[3],
 P.~Petrov,
 A.~Shaikhiev$\,$\footnotemark[4],
 R.~Volpe$\,$\footnotemark[5]\\[2mm]

{\bf TRIUMF, Vancouver, British Columbia, Canada}\\
 T.~Numao,
 Y.~Petrov,
 B.~Velghe\\[2mm]

{\bf University of British Columbia, Vancouver, British Columbia, Canada}\\
 D.~Bryman$\,$\footnotemark[6],
 J.~Fu\\[2mm]

{\bf Charles University, Prague, Czech Republic}\\
 T.~Husek$\,$\footnotemark[7],
 J.~Jerhot$\,$\footnotemark[8],
 K.~Kampf,
 M.~Zamkovsky\\[2mm]

{\bf Institut f\"ur Physik and PRISMA Cluster of Excellence, Universit\"at Mainz, Mainz, Germany}\\
 R.~Aliberti$\,$\footnotemark[9],
 G.~Khoriauli$\,$\footnotemark[10],
 J.~Kunze,
 D.~Lomidze$\,$\footnotemark[11],
 L.~Peruzzo,
 M.~Vormstein,
 R.~Wanke\\[2mm]

{\bf Dipartimento di Fisica e Scienze della Terra dell'Universit\`a e INFN, Sezione di Ferrara, Ferrara, Italy}\\
 P.~Dalpiaz,
 M.~Fiorini,
 I.~Neri,
 A.~Norton$\,$\footnotemark[12],
 F.~Petrucci,
 H.~Wahl$\,$\footnotemark[13]\\[2mm]

{\bf INFN, Sezione di Ferrara, Ferrara, Italy}\\
 A.~Cotta Ramusino,
 A.~Gianoli\\[2mm]

{\bf Dipartimento di Fisica e Astronomia dell'Universit\`a e INFN, Sezione di Firenze, Sesto Fiorentino, Italy}\\
 E.~Iacopini,
 G.~Latino,
 M.~Lenti,
 A.~Parenti\\[2mm]

{\bf INFN, Sezione di Firenze, Sesto Fiorentino, Italy}\\
 A.~Bizzeti$\,$\footnotemark[14],
 F.~Bucci\\[2mm]

{\bf Laboratori Nazionali di Frascati, Frascati, Italy}\\
 A.~Antonelli,
 G.~Georgiev$\,$\footnotemark[15],
 V.~Kozhuharov$\,$\footnotemark[15],
 G.~Lanfranchi,
 S.~Martellotti,
 M.~Moulson,
 T.~Spadaro\\[2mm]

{\bf Dipartimento di Fisica ``Ettore Pancini'' e INFN, Sezione di Napoli, Napoli, Italy}\\
 F.~Ambrosino,
 T.~Capussela,
 M.~Corvino$\,$\footnotemark[1],
 D.~Di Filippo,
 P.~Massarotti,
 M.~Mirra,
 M.~Napolitano,
 G.~Saracino\\[2mm]

{\bf Dipartimento di Fisica e Geologia dell'Universit\`a e INFN, Sezione di Perugia, Perugia, Italy}\\
 G.~Anzivino,
 F.~Brizioli,
 E.~Imbergamo,
 R.~Lollini,
 R.~Piandani$\,$\footnotemark[16],
 C.~Santoni\\[2mm]

{\bf INFN, Sezione di Perugia, Perugia, Italy}\\
 M.~Barbanera,
 P.~Cenci,
 B.~Checcucci,
 P.~Lubrano,
 M.~Lupi$\,$\footnotemark[17],
 M.~Pepe,
 M.~Piccini\\[2mm]

{\bf Dipartimento di Fisica dell'Universit\`a e INFN, Sezione di Pisa, Pisa, Italy}\\
 F.~Costantini,
 L.~Di Lella$\,$\footnotemark[13],
 N.~Doble$\,$\footnotemark[13],
 M.~Giorgi,
 S.~Giudici,
 G.~Lamanna,
 E.~Lari,
 E.~Pedreschi,
 M.~Sozzi\\[2mm]

{\bf INFN, Sezione di Pisa, Pisa, Italy}\\
 C.~Cerri,
 R.~Fantechi,
 L.~Pontisso,
 F.~Spinella\\[2mm]

{\bf Scuola Normale Superiore e INFN, Sezione di Pisa, Pisa, Italy}\\
 I.~Mannelli\\[2mm]

{\bf Dipartimento di Fisica, Sapienza Universit\`a di Roma e INFN, Sezione di Roma I, Roma, Italy}\\
 G.~D'Agostini,
 M.~Raggi\\[2mm]

{\bf INFN, Sezione di Roma I, Roma, Italy}\\
 A.~Biagioni,
 E.~Leonardi,
 A.~Lonardo,
 P.~Valente,
 P.~Vicini\\[2mm]

{\bf INFN, Sezione di Roma Tor Vergata, Roma, Italy}\\
 R.~Ammendola,
 V.~Bonaiuto$\,$\footnotemark[18],
 A.~Fucci,
 A.~Salamon,
 F.~Sargeni$\,$\footnotemark[19]\\[2mm]

{\bf Dipartimento di Fisica dell'Universit\`a e INFN, Sezione di Torino, Torino, Italy}\\
 R.~Arcidiacono$\,$\footnotemark[20],
 B.~Bloch-Devaux,
 M.~Boretto$\,$\footnotemark[1],
 E.~Menichetti,
 E.~Migliore,
 D.~Soldi\\[2mm]

{\bf INFN, Sezione di Torino, Torino, Italy}\\
 C.~Biino,
 A.~Filippi,
 F.~Marchetto\\[2mm]

{\bf Instituto de F\'isica, Universidad Aut\'onoma de San Luis Potos\'i, San Luis Potos\'i, Mexico}\\
 J.~Engelfried,
 N.~Estrada-Tristan$\,$\footnotemark[21]\\[2mm]

{\bf Horia Hulubei National Institute of Physics for R\&D in Physics and Nuclear Engineering, Bucharest-Magurele, Romania}\\
 A. M.~Bragadireanu,
 S. A.~Ghinescu,
 O. E.~Hutanu\\[2mm]

{\bf Joint Institute for Nuclear Research, Dubna, Russia}\\
 A.~Baeva,
 D.~Baigarashev,
 D.~Emelyanov,
 T.~Enik,
 V.~Falaleev,
 V.~Kekelidze,
 A.~Korotkova,
 L.~Litov$\,$\footnotemark[15],
 D.~Madigozhin,
 M.~Misheva$\,$\footnotemark[22],
 N.~Molokanova,
 S.~Movchan,
 I.~Polenkevich,
 Yu.~Potrebenikov,
 S.~Shkarovskiy,
 A.~Zinchenko$\,$\renewcommand{\thefootnote}{\fnsymbol{footnote}}\footnotemark[2]\renewcommand{\thefootnote}{\arabic{footnote}}\\[2mm]

{\bf Institute for Nuclear Research of the Russian Academy of Sciences, Moscow, Russia}\\
 S.~Fedotov,
 E.~Gushchin,
 A.~Khotyantsev,
 Y.~Kudenko$\,$\footnotemark[23],
 V.~Kurochka,
 M.~Medvedeva,
 A.~Mefodev\\[2mm]

{\bf Institute for High Energy Physics - State Research Center of Russian Federation, Protvino, Russia}\\
 S.~Kholodenko,
 V.~Kurshetsov,
 V.~Obraztsov,
 A.~Ostankov$\,$\renewcommand{\thefootnote}{\fnsymbol{footnote}}\footnotemark[2]\renewcommand{\thefootnote}{\arabic{footnote}},
 V.~Semenov$\,$\renewcommand{\thefootnote}{\fnsymbol{footnote}}\footnotemark[2]\renewcommand{\thefootnote}{\arabic{footnote}},
 V.~Sugonyaev,
 O.~Yushchenko\\[2mm]

{\bf Faculty of Mathematics, Physics and Informatics, Comenius University, Bratislava, Slovakia}\\
 L.~Bician$\,$\footnotemark[1],
 T.~Blazek,
 V.~Cerny,
 Z.~Kucerova\\[2mm]

{\bf CERN,  European Organization for Nuclear Research, Geneva, Switzerland}\\
 J.~Bernhard,
 A.~Ceccucci,
 H.~Danielsson,
 N.~De Simone$\,$\footnotemark[24],
 F.~Duval,
 B.~D\"obrich,
 L.~Federici,
 E.~Gamberini,
 L.~Gatignon,
 R.~Guida,
 F.~Hahn$\,$\renewcommand{\thefootnote}{\fnsymbol{footnote}}\footnotemark[2]\renewcommand{\thefootnote}{\arabic{footnote}},
 E. B.~Holzer,
 B.~Jenninger,
 M.~Koval$\,$\footnotemark[25],
 P.~Laycock$\,$\footnotemark[3],
 G.~Lehmann Miotto,
 P.~Lichard,
 A.~Mapelli,
 R.~Marchevski,
 K.~Massri,
 M.~Noy,
 V.~Palladino$\,$\footnotemark[26],
 M.~Perrin-Terrin$\,$\footnotemark[27]$^,\,$\footnotemark[28],
 J.~Pinzino$\,$\footnotemark[29],
 V.~Ryjov,
 S.~Schuchmann$\,$\footnotemark[13],
 S.~Venditti\\[2mm]

{\bf University of Birmingham, Birmingham, United Kingdom}\\
 T.~Bache,
 M. B.~Brunetti$\,$\footnotemark[30],
 V.~Duk$\,$\footnotemark[31],
 V.~Fascianelli$\,$\footnotemark[32],
 J. R.~Fry,
 F.~Gonnella,
 E.~Goudzovski$\renewcommand{\thefootnote}{\fnsymbol{footnote}}\footnotemark[1]\renewcommand{\thefootnote}{\arabic{footnote}}$,
 J.~Henshaw,
 L.~Iacobuzio,
 C.~Lazzeroni,
 N.~Lurkin$\,$\footnotemark[8],
 F.~Newson,
 C.~Parkinson$\,$\footnotemark[8],
 A.~Romano,
 A.~Sergi$\,$\footnotemark[33],
 A.~Sturgess,
 J.~Swallow\\[2mm]
\newpage
{\bf University of Bristol, Bristol, United Kingdom}\\
 H.~Heath,
 R.~Page,
 S.~Trilov\\[2mm]

{\bf University of Glasgow, Glasgow, United Kingdom}\\
 B.~Angelucci,
 D.~Britton,
 C.~Graham,
 D.~Protopopescu\\[2mm]

{\bf University of Lancaster, Lancaster, United Kingdom}\\
 J.~Carmignani,
 J. B.~Dainton,
 R. W. L.~Jones,
 G.~Ruggiero\\[2mm]

{\bf University of Liverpool, Liverpool, United Kingdom}\\
 L.~Fulton,
 D.~Hutchcroft,
 E.~Maurice$\,$\footnotemark[34],
 B.~Wrona\\[2mm]

{\bf George Mason University, Fairfax, Virginia, USA}\\
 A.~Conovaloff,
 P.~Cooper,
 D.~Coward$\,$\footnotemark[35],
 P.~Rubin\\[2mm]

\end{raggedright}
%
%
\setcounter{footnote}{0}
\renewcommand{\thefootnote}{\fnsymbol{footnote}}
\footnotetext[1]{Corresponding author: Evgueni Goudzovski, email: evgueni.goudzovski@cern.ch}
\footnotetext[2]{Deceased}
\renewcommand{\thefootnote}{\arabic{footnote}}

\footnotetext[1]{Present address: CERN,  European Organization for Nuclear Research, CH-1211 Geneva 23, Switzerland}
\footnotetext[2]{Also at Laboratori Nazionali di Frascati, I-00044 Frascati, Italy}
\footnotetext[3]{Present address: Brookhaven National Laboratory, Upton, NY 11973, USA}
\footnotetext[4]{Also at Institute for Nuclear Research of the Russian Academy of Sciences, 117312 Moscow, Russia}
\footnotetext[5]{Present address: Faculty of Mathematics, Physics and Informatics, Comenius University, 842 48, Bratislava, Slovakia}
\footnotetext[6]{Also at TRIUMF, Vancouver, British Columbia, V6T 2A3, Canada}
\footnotetext[7]{Present address: Department of Astronomy and Theoretical Physics, Lund University, Lund, SE 223-62, Sweden}
\footnotetext[8]{Present address: Universit\'e Catholique de Louvain, B-1348 Louvain-La-Neuve, Belgium}
\footnotetext[9]{Present address: Institut f\"ur Kernphysik and Helmholtz Institute Mainz, Universit\"at Mainz, Mainz, D-55099, Germany}
\footnotetext[10]{Present address: Universit\"at W\"urzburg, D-97070 W\"urzburg, Germany}
\footnotetext[11]{Present address: European XFEL GmbH, D-22761 Hamburg, Germany}
\footnotetext[12]{Present address: University of Glasgow, Glasgow, G12 8QQ, UK}
\footnotetext[13]{Present address: Institut f\"ur Physik and PRISMA Cluster of excellence, Universit\"at Mainz, D-55099 Mainz, Germany}
\footnotetext[14]{Also at Dipartimento di Fisica, Universit\`a di Modena e Reggio Emilia, I-41125 Modena, Italy}
\footnotetext[15]{Also at Faculty of Physics, University of Sofia, BG-1164 Sofia, Bulgaria}
\footnotetext[16]{Present address: Institut f\"ur Experimentelle Teilchenphysik (KIT), D-76131 Karlsruhe, Germany}
\footnotetext[17]{Present address: Institut am Fachbereich Informatik und Mathematik, Goethe Universit\"at, D-60323 Frankfurt am Main, Germany}
\footnotetext[18]{Also at Department of Industrial Engineering, University of Roma Tor Vergata, I-00173 Roma, Italy}
\footnotetext[19]{Also at Department of Electronic Engineering, University of Roma Tor Vergata, I-00173 Roma, Italy}
\footnotetext[20]{Also at Universit\`a degli Studi del Piemonte Orientale, I-13100 Vercelli, Italy}
\footnotetext[21]{Also at Universidad de Guanajuato, Guanajuato, Mexico}
\footnotetext[22]{Present address: Institute of Nuclear Research and Nuclear Energy of Bulgarian Academy of Science (INRNE-BAS), BG-1784 Sofia, Bulgaria}
\footnotetext[23]{Also at National Research Nuclear University (MEPhI), 115409 Moscow and Moscow Institute of Physics and Technology, 141701 Moscow region, Moscow, Russia}
\footnotetext[24]{Present address: DESY, D-15738 Zeuthen, Germany}
\footnotetext[25]{Present address: Charles University, 116 36 Prague 1, Czech Republic}
\footnotetext[26]{Present address: Physics Department, Imperial College London, London, SW7 2BW, UK}
\footnotetext[27]{Present address: Aix Marseille University, CNRS/IN2P3, CPPM, F-13288, Marseille, France}
\footnotetext[28]{Also at Universit\'e Catholique de Louvain, B-1348 Louvain-La-Neuve, Belgium}
\footnotetext[29]{Present address: INFN, Sezione di Pisa, I-56100 Pisa, Italy}
\footnotetext[30]{Present address: Department of Physics, University of Warwick, Coventry, CV4 7AL, UK}
\footnotetext[31]{Present address: INFN, Sezione di Perugia, I-06100 Perugia, Italy}
\footnotetext[32]{Present address: Center for theoretical neuroscience, Columbia University, New York, NY 10027, USA}
\footnotetext[33]{Present address: Dipartimento di Fisica dell'Universit\`a e INFN, Sezione di Genova, I-16146 Genova, Italy}
\footnotetext[34]{Present address: Laboratoire Leprince Ringuet, F-91120 Palaiseau, France}
\footnotetext[35]{Also at SLAC National Accelerator Laboratory, Stanford University, Menlo Park, CA 94025, USA}

\newpage


\section*{Introduction}

All Standard Model (SM) fermions except neutrinos are known to exhibit both chiralities. The existence of right-handed neutrinos, or heavy neutral leptons (HNLs), is hypothesised in many SM extensions to generate non-zero masses of the SM neutrinos via the seesaw mechanism~\cite{pbc19}. For example, the Neutrino Minimal Standard Model~\cite{nuMSM} accounts for dark matter, baryogenesis, neutrino masses and oscillations by postulating two HNLs in the MeV--GeV mass range and a third HNL at the keV mass scale, which is a dark matter candidate.

Mixing between HNLs (denoted $N$ below) and active neutrinos gives rise to HNL production in meson decays. The expected branching fraction of the $K^+\to\mu^+N$ decay is~\cite{sh80}
\begin{displaymath}
{\cal B}(K^+\to\mu^+ N) = {\cal B}(K^+\to\mu^+\nu) \cdot \rho_\mu(m_N) \cdot |U_{\mu 4}|^2,
\end{displaymath}
where ${\cal B}(K^+\to\mu^+\nu)$ is the measured branching fraction of the SM leptonic decay~\cite{pdg}, $|U_{\mu 4}|^2$ is the mixing parameter, and $\rho_\mu(m_N)$ is a kinematic factor which depends on the HNL mass $m_N$:
\begin{equation}
\rho_\mu(m_N) = \frac {(x+y)-(x-y)^2} {x(1-x)^2} \cdot \lambda^{1/2}(1,x,y),
\label{eq:rho}
\end{equation}
with $x=(m_\mu/m_K)^2$, $y=(m_N/m_K)^2$ and $\lambda(1,x,y)=1+x^2+y^2-2(x+y+xy)$.
The factor $\rho_\mu(m_N)$ increases from unity at $m_N=0$ to a maximum of 4.13 at $m_N=263$~MeV/$c^2$, and decreases to zero at the kinematic limit $m_N=m_K-m_\mu$. Assuming that the HNL decays exclusively to SM particles, its lifetime in the mass range $m_N<m_K$ exceeds $10^{-4}/|U_4|^2~\mu$s, where $|U_4|^2$ is the largest of the three coupling parameters $|U_{\ell 4}|^2$ ($\ell=e,\mu,\tau$)~\cite{bo18}. Therefore under the above assumption, and additionally assuming conservatively that $|U_{\ell4}|^2<10^{-4}$, the HNL can be considered stable in production-search experiments.

A new light gauge boson has been proposed as an explanation to the muon $g-2$ anomaly~\cite{gn01}. A particular scenario, which also accommodates dark matter (DM) freeze-out, involves a scalar or vector hidden sector mediator $X$ coupling preferentially to the muon. This mediator is expected to be produced in $K^+\to\mu^+\nu X$ decays with an estimated branching fraction of ${\cal O}(10^{-8})$ in case $m_X<m_K-m_\mu$, and is expected to decay promptly with a sizeable invisible branching fraction~\cite{kr20}.
In the light DM freeze-out model, the $X\to\chi\bar\chi$ decay is expected, where $\chi$ is the DM particle.

The $K^+\to\mu^+\nu\nu\bar\nu$ decay occurs within the SM at second order in the Fermi constant $G_F$, and the expected branching fraction at leading order in chiral perturbation theory, ${\cal B_{\rm SM}}=1.62\times 10^{-16}$~\cite{go16}, is experimentally out of reach. The strongest upper limit to date, ${\cal B} (K^+\to\mu^+\nu\nu\bar\nu)<2.4\times 10^{-6}$ at~90\%~CL, has been established by the BNL-E949 experiment~\cite{ar16}.

The $K^+\to\mu^+N$, $K^+\to\mu^+\nu X$ and $K^+\to\mu^+\nu\nu\bar\nu$ decays with invisible $N$ and $X$ particles are characterised by a single muon and missing energy in the final state. Searches for these decays using the data collected by the NA62 experiment at CERN in 2016--2018 are reported here. The $N$ particle is interpreted as a HNL, and the results are presented as upper limits of the extended neutrino mixing matrix element $|U_{\mu4}|^2$ for $m_N$ in the range 200--384~MeV/$c^2$, with the assumption that the HNL lifetime exceeds 50 ns. For the $K^+\to\mu^+\nu X$ decays (in a number of $m_X$ hypotheses within the range 10--370~MeV/$c^2$) and the $K^+\to\mu^+\nu\nu\bar\nu$ decay, upper limits on the branching fractions are reported.


\section{Beam, detector and data sample}
\label{sec:detector}

\begin{figure}[t]
\begin{center}
\resizebox{\textwidth}{!}{\includegraphics{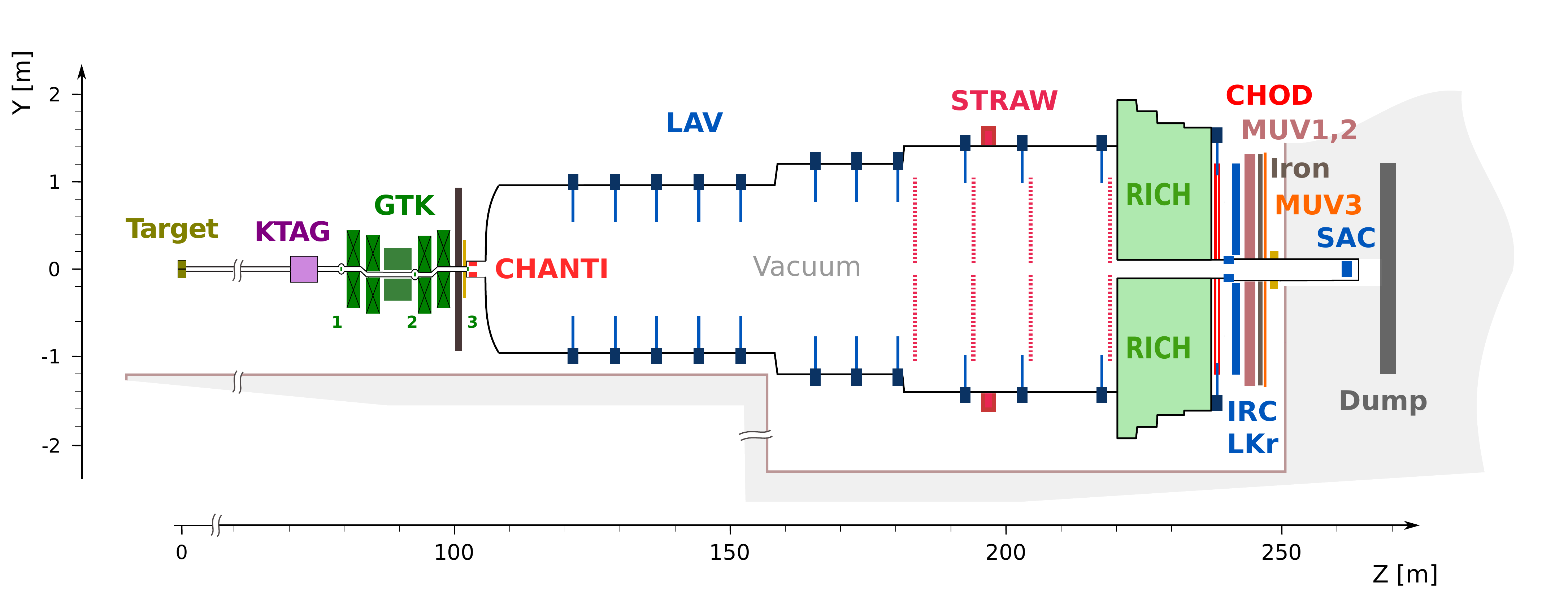}}
\put(-343,67){\scriptsize\color{burgundy}\rotatebox{90}{\textbf{\textsf{SCR}}}}
\put(-325,121){\scriptsize\color{burgundy}\rotatebox{90}{\textbf{\textsf{COL}}}}
\put(-172,46){\scriptsize\color{red}{\textbf{\textsf{M}}}}
\end{center}
\vspace{-15mm}
\caption{Schematic side view of the NA62 beamline and detector.}
\label{fig:detector}
\end{figure}

The layout of the NA62 beamline and detector~\cite{na62-detector} is shown schematically in Fig.~\ref{fig:detector}. An unseparated secondary beam of $\pi^+$ (70\%), protons (23\%) and $K^+$ (6\%) is created by directing 400~GeV/$c$ protons extracted from the CERN SPS onto a beryllium target in spills of 3~s effective duration. The central beam momentum is 75~GeV/$c$, with a momentum spread of 1\% (rms).

Beam kaons are tagged with 70~ps time resolution by a differential Cherenkov counter (KTAG) using as radiator nitrogen gas at 1.75~bar pressure contained in a 5~m long vessel. Beam particle positions, momenta and times (to better than 100~ps resolution) are measured by a silicon pixel spectrometer consisting of three stations (GTK1,2,3) and four dipole magnets. A muon scraper (SCR) is installed between GTK1 and GTK2. A 1.2~m thick steel collimator (COL) with a central aperture of $76\times40$~mm$^2$ and outer dimensions of $1.7\times1.8$~m$^2$ is placed upstream of GTK3 to absorb hadrons from upstream $K^+$ decays (a variable aperture collimator of $0.15\times0.15$~m$^2$ outer dimensions was used up to early 2018). Inelastic interactions of beam particles in GTK3 are detected by an array of scintillator hodoscopes (CHANTI). The beam is delivered into a vacuum tank evacuated to a pressure of $10^{-6}$~mbar, which contains a 75~m long fiducial decay volume (FV) starting 2.6~m downstream of GTK3. The beam divergence at the FV entrance is 0.11~mrad (rms) in both horizontal and vertical planes. Downstream of the FV, undecayed beam particles continue their path in vacuum.

Momenta of charged particles produced by $K^+$ decays in the FV are measured by a magnetic spectrometer (STRAW) located in the vacuum tank downstream of the FV. The spectrometer consists of four tracking chambers made of straw tubes, and a dipole magnet (M) located between the second and third chambers that provides a horizontal momentum kick of 270~MeV/$c$. The momentum resolution achieved is $\sigma_p/p = (0.30\oplus 0.005 p)\%$, where the momentum $p$ is expressed in GeV/$c$.

A ring-imaging Cherenkov detector (RICH), consisting of a 17.5~m long vessel filled with neon at atmospheric pressure (with a Cherenkov threshold for muons of 9.5~GeV/$c$), is used for the identification of charged particles and for time measurement with 70~ps precision for particles well above the threshold. Two scintillator hodoscopes (CHOD), which include a matrix of tiles and two planes of slabs arranged in four quadrants downstream of the RICH, provide trigger signals and time measurements with 200~ps precision.

A $27X_0$ thick quasi-homogeneous liquid krypton (LKr) electromagnetic calorimeter is used for particle identification and photon detection. The calorimeter has an active volume of 7~m$^3$, is segmented in the transverse direction into 13248 projective cells of approximately $2\!\times\!2$~cm$^2$, and provides an energy resolution $\sigma_E/E=(4.8/\sqrt{E}\oplus11/E\oplus0.9)\%$, where $E$ is expressed in GeV. To achieve hermetic acceptance for photons emitted in the FV by $K^+$ decays at angles up to 50~mrad to the beam axis, the LKr calorimeter is supplemented by annular lead glass detectors (LAV) installed in 12~positions inside and downstream of the vacuum tank, and two lead/scintillator sampling calorimeters (IRC, SAC) located close to the beam axis. An iron/scintillator sampling hadronic calorimeter formed of two modules (MUV1,2) and a muon detector (MUV3) consisting of 148~scintillator tiles located behind an 80~cm thick iron wall are used for particle identification.

%
%

The data sample used for this analysis is obtained from $0.92\times 10^6$ SPS spills recorded during 410 days of operation in 2016--2018, with the typical beam intensity increasing over time from \mbox{$1.3\times 10^{12}$} to \mbox{$2.2\times 10^{12}$} protons per spill. The latter value corresponds to a mean instantaneous beam particle rate at the FV entrance of 500~MHz, and a mean $K^+$ decay rate in the FV of 3.7~MHz. Data recorded with a minimum-bias trigger based on CHOD signals~\cite{am19}, downscaled by a factor of 400, is used for the analysis. This trigger is 99\% efficient for single charged particles in the CHOD acceptance.


\section{Measurement principles and event selection}
\label{sec:selection}

The rates of the signal processes are measured with respect to the $K^+\to\mu^+\nu$ decay rate. This approach benefits from first-order cancellations of residual detector  inefficiencies not fully accounted for in simulations, as well as trigger  inefficiencies and random veto losses common to signal and normalization modes.

Candidate signal decays, as well as the $K^+\to\mu^+\nu$ decay, are characterised by a single muon and no other detectable particles in the final state. Backgrounds are due to beam particle decays upstream of the vacuum tank, decays to multiple detectable particles, and inelastic interactions of beam particles in GTK3. Event selection is optimized to suppress these backgrounds. The principal selection criteria are listed below.
\begin{itemize}
\item A positively charged muon track is required to be reconstructed in the STRAW spectrometer with momentum in the range 5--70~GeV/$c$. The track's trajectory through the STRAW chambers and its extrapolation to the LKr calorimeter, CHOD and MUV3 should be within the geometrical acceptance of these detectors. The muon time is evaluated using the RICH and CHOD signals spatially associated with the track.
\item Particle identification criteria are applied to the STRAW track to suppress the backgrounds due to misidentification. The ratio of the energy deposited in the LKr calorimeter, $E$, to the momentum, $p$, measured by the STRAW spectrometer is required to be $E/p<0.2$. For tracks with momentum below 30~GeV/$c$, a particle identification algorithm is applied based on the RICH signal pattern within 3~ns of the CHOD time. In particular, tracks with momenta below the muon Cherenkov threshold must not be identified as positrons. At least one signal in the MUV3 detector must be within 3~ns of the muon time and spatially consistent with the projected track impact point in the MUV3 front plane.
\item Backgrounds from $K^+\to\mu^+\nu$ decays upstream of the KTAG and $\pi^+\to\mu^+\nu$ decays upstream of GTK3, in coincidence with a beam pion or proton track in the GTK, are suppressed by requiring a kaon signal in the KTAG detector within 1~ns of the muon time.
\item The decay vertex is defined as the point of closest approach of the $K^+$ track in the GTK and the muon track in the STRAW, taking into account the stray magnetic field in the vacuum tank. Identification of the $K^+$ track in the GTK relies on the time difference, $\Delta t_{\rm GK}$, between a GTK track and the KTAG signal, and spatial compatibility of the GTK and STRAW tracks quantified by the distance, $d$, of closest approach. A discriminant ${\cal D}(\Delta t_{\rm GK}, d)$ is defined using the $\Delta t_{\rm GK}$ and $d$ distributions measured with $K^+\to\pi^+\pi^+\pi^-$ decays~\cite{pinn20}. Among GTK tracks with $|\Delta t_{\rm GK}|<0.5$~ns, the track of the parent kaon is assumed to be the one with the ${\cal D}$ value most consistent with a $K^+\to\mu^+$ decay. It is required that $d<7$~mm to reduce the background from upstream decays.
\item Background from $K^+\to\mu^+\nu$ decays between KTAG and GTK3 with pileup in the GTK is suppressed by geometrical conditions. The reconstructed $K^+$ decay vertex is required to be located in the FV at a minimum distance from the start of the FV, varying from 8~m to 35~m depending on the angle between the $K^+$ momentum in the laboratory frame and the muon momentum in the $K^+$ rest frame.
\item Backgrounds from $K^+$ decays to multiple detectable particles are suppressed by veto conditions. The muon track must not form a vertex with any additional STRAW track segment. Energy deposits are not allowed in the LKr calorimeter that are spatially incompatible with the muon track within 12~ns of the muon time. No activity is allowed in the large-angle (LAV) or small-angle (SAC, IRC) photon veto detectors within 3~ns of the muon time, or in the CHANTI detector within 4~ns of the muon time. No more than two signals in the CHOD tiles within 6~ns of the muon time, and no more than three signals in the RICH PMTs within 3~ns of the muon time, spatially incompatible with the muon track, are allowed. Data loss due to the veto conditions from accidental activity (random veto) averaged over the data sample is measured to be about 30\%.
\end{itemize}

The squared missing mass is computed as $m_{\rm miss}^2=(P_K-P_\mu)^2$, where $P_K$ and $P_\mu$ are the kaon and muon 4-momenta, obtained from the 3-momenta measured by the GTK and STRAW spectrometers under the $K^+$ and $\mu^+$ mass hypotheses.

\begin{figure}[t]
\renewcommand{\arraystretch}{0}
\centering
\setlength\tabcolsep{0pt}
\begin{tabular}[t]{cc}
\begin{subfigure}{0.5\textwidth}
\centering
\includegraphics[width=1.0\linewidth]{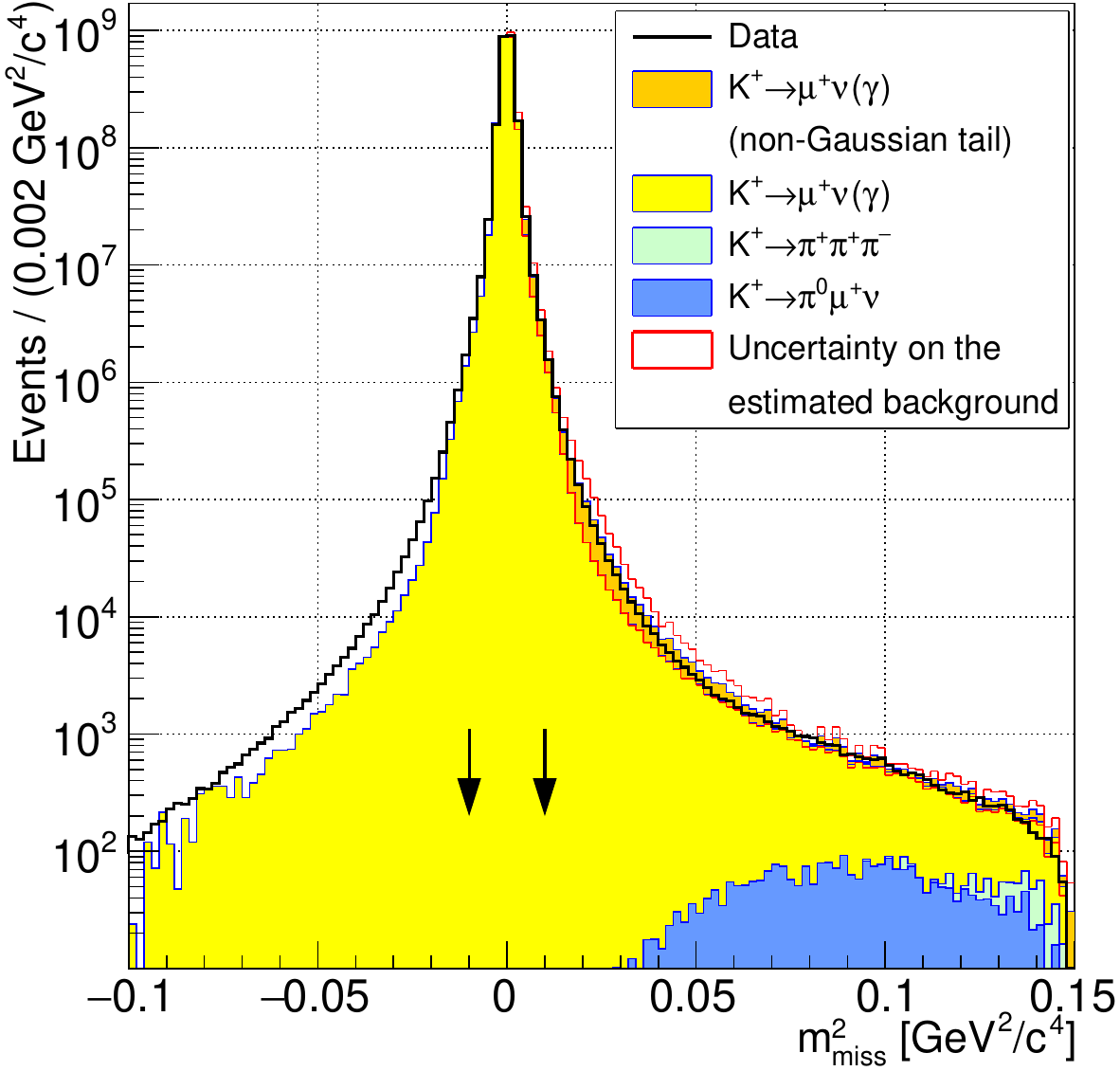}
\end{subfigure}
&
\begin{tabular}{c}%
\begin{subfigure}[t]{0.5\textwidth}
\centering
\includegraphics[width=1.0\textwidth]{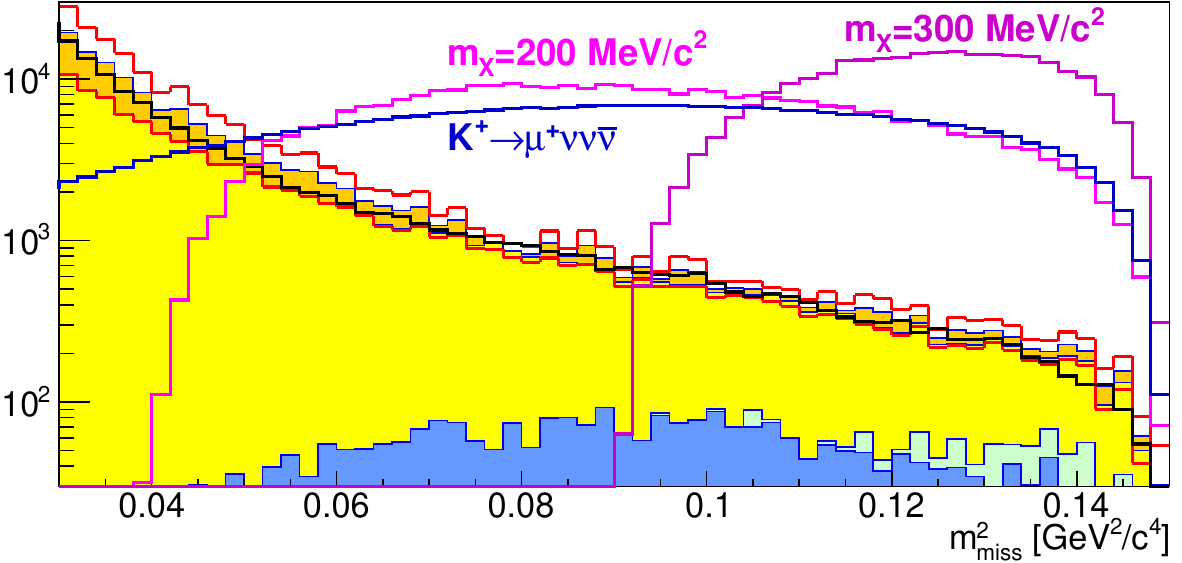}
\end{subfigure}\\
\begin{subfigure}[t]{0.5\textwidth}
\centering
\includegraphics[width=1.0\textwidth]{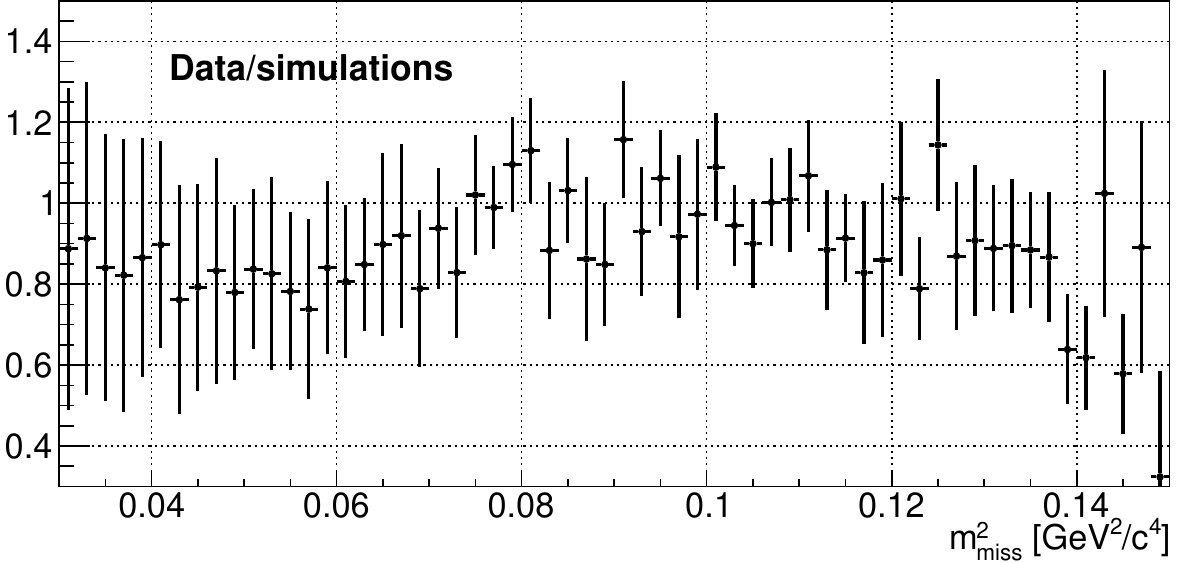}
\end{subfigure}
\end{tabular}\\
\end{tabular}
\caption{Left: reconstructed $m_{\rm miss}^2$ distributions for data and the estimated background. The full uncertainties ($\pm1\sigma$) in each mass bin of the background spectrum for $m_{\rm miss}^2>0$ are shown with a contour. The boundaries of the SM signal region \mbox{$|m_{\rm miss}^2|<0.01~{\rm GeV}^2/c^4$} used for normalisation are indicated with arrows. Top-right: the region $m_{\rm miss}^2>0.03~{\rm GeV}^2/c^4$, with simulated hypothetical $K^+\to\mu^+\nu X$ (scalar mediator model, two $m_X$ values) and $K^+\to\mu^+\nu\nu\bar\nu$ signals with branching fractions of $10^{-4}$. Bottom-right: ratio of data and simulated spectra in the region $m_{\rm miss}^2>0.03~{\rm GeV}^2/c^4$ with the full uncertainties. Systematic components of the uncertainties are correlated among the bins.}
\label{fig:mmiss2}
\end{figure}

Monte Carlo simulations of particle interactions with the detector and its response are performed with a software package based on the \geant toolkit~\cite{geant4}. The $m_{\rm miss}^2$ spectra of the selected events from data and simulated samples, and their ratio, are displayed in Fig.~\ref{fig:mmiss2}. The signal from the SM leptonic decay \mbox{$K^+\to\mu^+\nu$} is observed as a peak at $m_{\rm miss}^2=0$ with a resolution of \mbox{$1.5\times10^{-3}~{\rm GeV}^2/c^4$}, and the SM signal region is defined in terms of the reconstructed squared missing mass as \mbox{$|m_{\rm miss}^2|<0.01~{\rm GeV}^2/c^4$}. In contrast, the $K^+\to\mu^+N$, $K^+\to\mu^+\nu X$ and $K^+\to\mu^+\nu\nu\bar\nu$ decays are characterised by larger $m_{\rm miss}^2$ values.

\boldmath
\section{Normalisation to the $K^+\to\mu^+\nu$ decay}
\unboldmath
\label{sec:norm}

The effective number of $K^+$ decays in the FV, denoted $N_K$, is evaluated using the number of $K^+\to\mu^+\nu$ candidates reconstructed in the data sample. The quantity $N_K$ is not corrected for trigger inefficiency and random veto effects, which cancel between signal and normalisation
thus making the $N_K$ value specific to this analysis. The background in the SM signal region is negligible (Fig.~\ref{fig:mmiss2}). It is found that
\begin{displaymath}
N_K = \frac{N_{\rm SM}}{A_{\rm SM} \cdot {\cal B}(K^+\to\mu^+\nu)} = (1.14\pm0.02)\times 10^{10},
\end{displaymath}
where $N_{\rm SM}=2.19\times 10^{9}$ is the number of selected data events in the SM signal region, $A_{\rm SM}=0.302\pm0.005$ is the acceptance of the selection for the $K^+\to\mu^+\nu$ decay evaluated using simulations, and ${\cal B}(K^+\to\mu^+\nu)=0.6356\pm0.0011$ is the branching fraction of this decay~\cite{pdg}. The uncertainty of $A_{\rm SM}$, which dominates that of $N_K$, is mainly systematic due to the accuracy of the simulation, and is evaluated by variation of the selection criteria including the algorithm used for identification of the $K^+$ track in the GTK.


\section{Background evaluation with simulations}
\label{sec:bkg}

The main backgrounds to the potential signals at large $m_{\rm miss}^2$ values are due to the $K^+\to\mu^+\nu\gamma$, $K^+\to\pi^0\mu^+\nu$ ($\pi^0\to\gamma\gamma$) and $K^+\to\pi^+\pi^+\pi^-$ decays inside and upstream of the vacuum tank. Their contributions are estimated with simulations. The $K^+\to\mu^+\nu\gamma$ decay is simulated including inner bremsstrahlung (IB) and structure-dependent processes, and the interference between these processes~\cite{bi92}.



The $K^+\to\mu^+\nu\gamma$ and $K^+\to\pi^0\mu^+\nu$ backgrounds arise from the photon detection inefficiency in the hermetic NA62 photon veto system, and photon conversions in the STRAW and RICH detectors. Photon detection inefficiency is modelled for the simulated events using the LAV, LKr, IRC and SAC inefficiencies measured as functions of photon energy using a $K^+\to\pi^+\pi^0$ decay sample~\cite{pi0inv}.
To evaluate the systematic uncertainties in the background estimates, an alternative photon veto response model is used for the simulated events involving photon detector inefficiencies increased by one sigma of the measurements, and a conservative assumption that photons converting upstream of the STRAW spectrometer dipole magnet are not detected in the LAV, IRC and SAC systems. The latter assumption accounts for the different photon veto conditions used in this analysis with respect to those used for the inefficiency measurements~\cite{pi0inv}. The resulting systematic uncertainty of the estimated background comes mainly from the limited accuracy of the LAV inefficiency measurements. In particular, the LAV inefficiency is measured to be $(0.30\pm0.06)\%$ for photons in the 0.3--3~GeV energy range, which contains most photons from $K^+\to\mu^+\nu\gamma$ decays intercepting the LAV geometrical acceptance.

The accuracy of the description of the non-Gaussian $m_{\rm miss}^2$ tails of the $K^+\to\mu^+\nu(\gamma)$ decay is affected by the limited precision in the simulation of beam particle pileup
and inefficiency in the GTK. This leads to a deficit of simulated events in the negative tail of the $m_{\rm miss}^2$ distribution populated by the $K^+\to\mu^+\nu(\gamma)$ decays only (Fig.~\ref{fig:mmiss2}). For example, a 40\% deficit is observed in the region $m_{\rm miss}^2<-0.05~{\rm GeV}^2/c^4$. To account for the missing component in the positive tail, it is assumed that the non-Gaussian tails of the $m_{\rm miss}^2$ spectrum are left-right symmetrical. A ``tail'' component (shown separately in Fig.~\ref{fig:mmiss2}) is added to the estimated background in each $m_{\rm miss}^2$ bin in the region $m_{\rm miss}^2>0$ equal to the difference between the data and simulated spectra in the symmetric mass bin with respect to $m_{\rm miss}^2=0$. A 100\% uncertainty is conservatively assigned to this component to account for the above assumption.

\newpage

The composition of the estimated background in the kinematic region $m_{\rm miss}^2>0.1~{\rm GeV}^2/c^4$ is reported in Table~\ref{tab:bkg_km2}. The largest component is the radiative $K^+\to\mu^+\nu\gamma$ (IB) tail, and its uncertainty is dominated by a contribution due to the accuracy of the description of the non-Gaussian tail. Further systematic uncertainties due to beam tuning, calibrations, trigger and reconstruction efficiency are negligible compared with the overall systematic uncertainty from the sources considered. The background represents an ${\cal O}(10^{-6})$ fraction of the number of reconstructed SM $K^+\to\mu^+\nu$ candidates. Within the region $m_{\rm miss}^2>0.03~{\rm GeV}^2/c^4$, the estimated background agrees with the data within uncertainties (taking into account bin-to-bin correlations of the systematic uncertainties) as shown in Fig.~\ref{fig:mmiss2}.

\begin{table}[t]
\caption{Estimated backgrounds in the kinematic region $m_{\rm miss}^2>0.1~{\rm GeV}^2/c^4$ with their uncertainties. The uncertainties
labelled ``PV'' are systematic due to the accuracy of the photon veto efficiency modelling (positively correlated among the background sources),
and the one labelled ``tail'' is systematic and accounts for the accuracy of the non-Gaussian $m_{\rm miss}^2$ tail simulation.}
\begin{center}
\vspace{-14mm}
\begin{tabular}{lrcrcrcl}
\hline
Background source & \multicolumn{7}{c}{Estimated background} \\
\hline
$K^+\to\mu^+\nu\gamma$ & 6224 & $\!\!\pm\!\!$ & $105_{\rm stat}$ & $\!\!\pm\!\!$ & $333_{\rm PV}$ & $\!\!\pm\!\!$ & $780_{\rm tail}$ \\
$K^+\to\pi^0\mu^+\nu$ & 1016 & $\!\!\pm\!\!$ & $47_{\rm stat}$ & $\!\!\pm\!\!$ & $178_{\rm PV}$\\
$K^+\to\pi^+\pi^+\pi^-$ & 309 & $\!\!\pm\!\!$ & $32_{\rm stat}$\\
\hline
Total background~~~~ & 7549 & $\!\!\pm\!\!$ & $119_{\rm stat}$ & $\!\!\pm\!\!$ & \multicolumn{3}{c}{$920_{\rm syst}$}\\
\hline
\end{tabular}
\end{center}
\vspace{-13mm}
\label{tab:bkg_km2}
\end{table}


\section{\boldmath Search for $K^+\to\mu^+N$ decays}
\label{sec:kmn}
\vspace{-1mm}

\begin{figure}[t]
\begin{center}
\resizebox{0.50\textwidth}{!}{\includegraphics{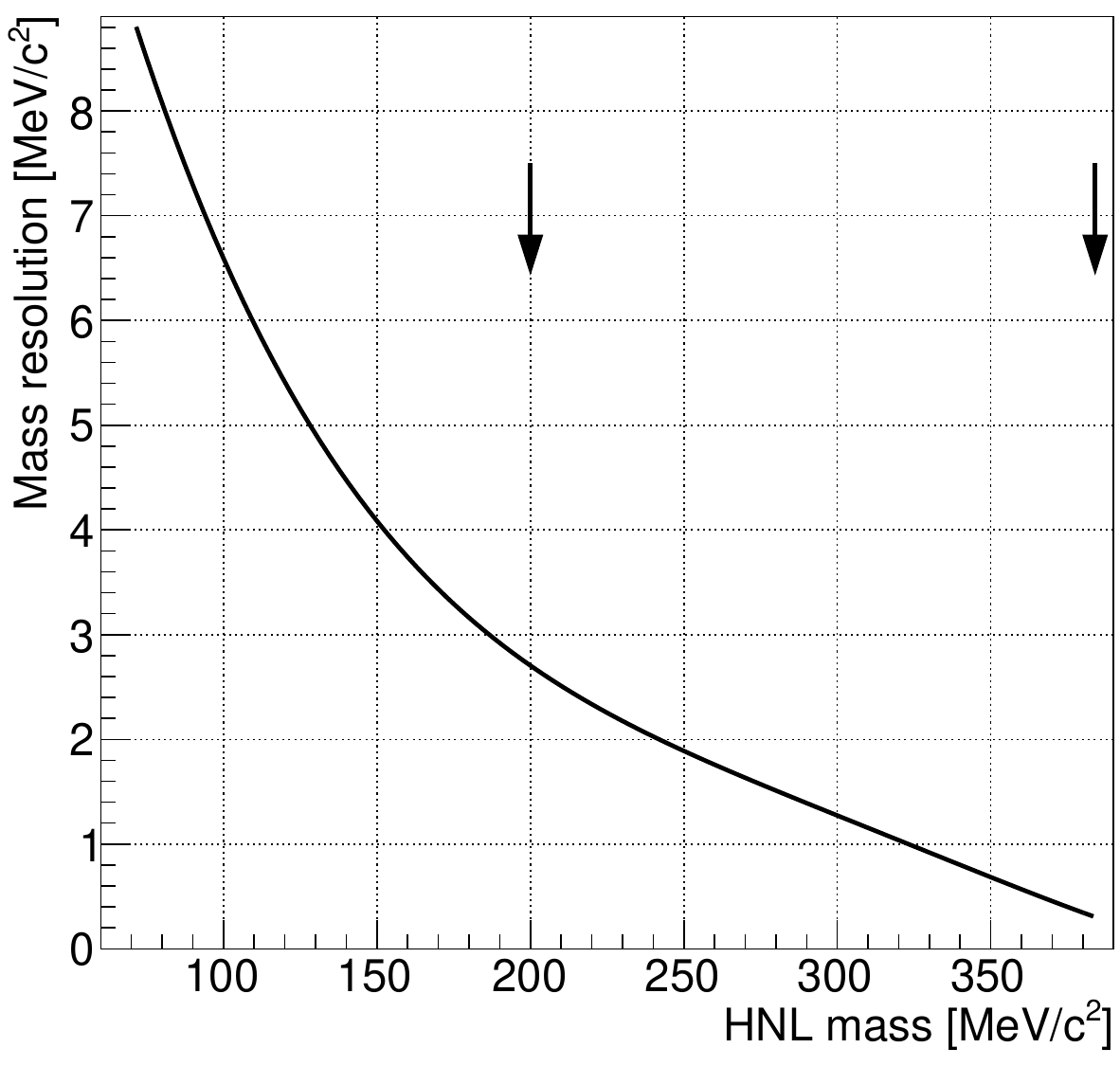}}%
\resizebox{0.50\textwidth}{!}{\includegraphics{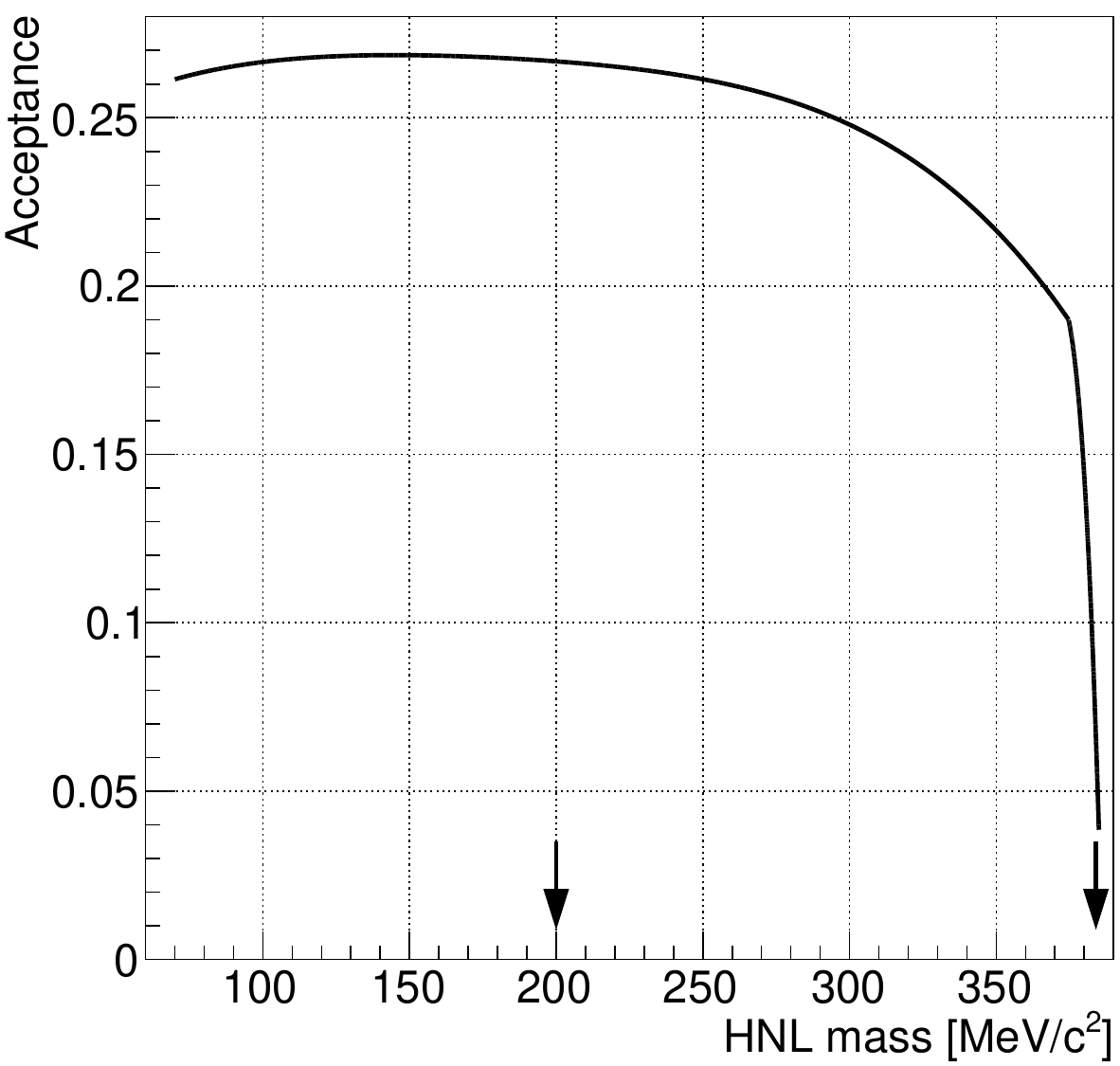}}
\end{center}
\vspace{-14mm}
\caption{HNL mass resolution $\sigma_m$ (left) and acceptance $A_N$ of the selection (right) evaluated from simulations as functions of the HNL mass. Boundaries of the HNL search region are indicated by vertical arrows.}
\vspace{-2mm}
\label{fig:resolution-acceptance}
\end{figure}

The $K^+\to\mu^+N$ process is investigated in 269 mass hypotheses, $m_N$, within the HNL search region 200--384~MeV/$c^2$. Distances between adjacent $m_N$ values considered are 1~(0.5)~MeV/$c^2$ below (above) the mass of 300~MeV/$c^2$. The decay is characterised by a narrow peak in the reconstructed missing mass ($m_{\rm miss}$) spectrum. Therefore the $K^+\to\mu^+N$ event selection requires that $|m_{\rm miss}-m_N|<1.5\sigma_m$ for each mass hypothesis $m_N$, where $\sigma_m$ is the mass resolution evaluated with simulations, as shown in Fig.~\ref{fig:resolution-acceptance}~(left). The resolution improves by a factor of three with respect to the NA62 2015 data sample collected without the GTK spectrometer~\cite{co18}.

Considering the peaking nature of the $K^+\to\mu^+N$ signal,
the background in each $m_N$ hypothesis is evaluated using sidebands in the reconstructed $m_{\rm miss}$ spectrum of the data events. This method is more precise than one based on simulation. Sidebands are defined in each mass hypothesis as $1.5\sigma_m<|m_{\rm miss}-m_N|<11.25\sigma_m$, additionally requiring that $m_{\rm miss}$ is within the range 188--386~MeV/$c^2$. The number of expected background events, $N_{\rm exp}$, within the $\pm1.5\sigma_m$ signal window is evaluated with a second-order polynomial fit to the sideband data of the $m_{\rm miss}$ spectrum, where the bin size is $0.75\sigma_m$. The uncertainty, $\delta N_{\rm exp}$, in the number of expected background events includes statistical and systematic components. The former comes from the uncertainties in the fit parameters, while the latter is evaluated as the difference between values of $N_{\rm exp}$ obtained from fits using second and third order polynomials. The dominant contribution to $\delta N_{\rm exp}$ is statistical, although systematic uncertainties become comparable as $m_N$ approaches the boundaries of the HNL search region. Systematic errors due to possible HNL signals in the sidebands are found to be negligible; this check is made assuming $|U_{\mu4}|^2$ to be equal to the expected sensitivity of the analysis. The uncertainty in the background estimate, $\delta N_{\rm exp}/N_{\rm exp}$, increases from 1--2\% for $m_N$ below 300~MeV/$c^2$ to 10\% at the upper limit of the HNL search region.


The signal selection acceptance, $A_N$, as a function of $m_N$ obtained with simulations assuming infinite HNL lifetime is displayed in Fig.~\ref{fig:resolution-acceptance}~(right). The acceptance for a mean lifetime of 50~ns (considering decays to detectable particles) is lower by ${\cal O}(1\%)$ in relative terms, making the results of the search valid for lifetimes in excess of 50~ns.  For shorter lifetimes, the HNL mean decay length in the laboratory frame becomes comparable to or smaller than the length of the apparatus. Acceptances for lifetimes of 5~(1)~ns decrease by factors up to 2~(10), depending on $m_N$. Simulations reproduce the $m_{\rm miss}^2$ resolution at the $K^+\to\mu^+\nu$ peak to a 1\% relative precision. Modelling of the resolution outside the peak is validated using data and simulated $K^+\to\pi^+\pi^+\pi^-$ decay samples; the corresponding systematic effects on $A_N$ do not exceed 2\% in relative terms~\cite{hnl-e}.

\begin{figure}[t]
\vspace{-2mm}
\begin{center}
\resizebox{0.50\textwidth}{!}{\includegraphics{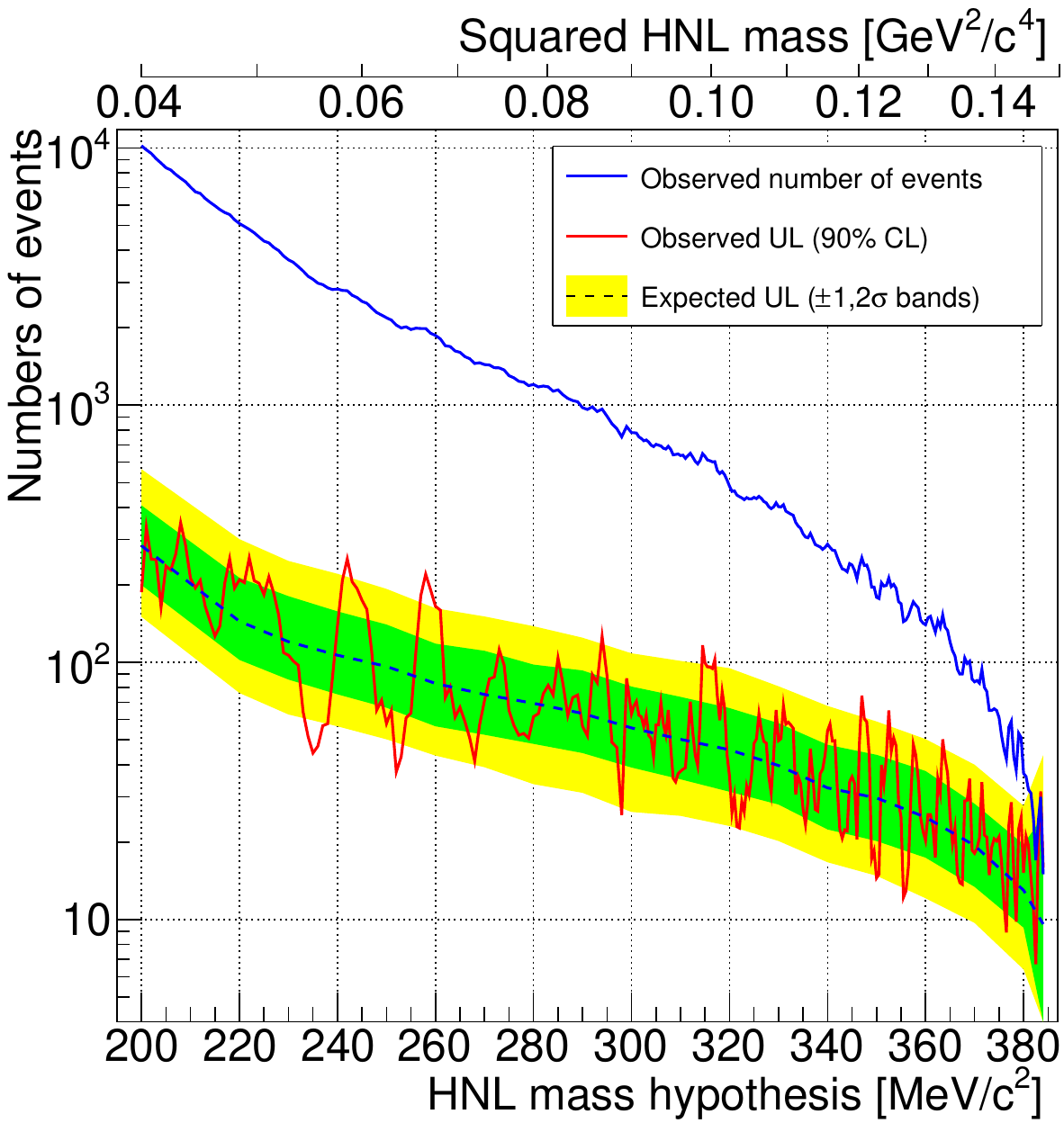}}%
\resizebox{0.50\textwidth}{!}{\includegraphics{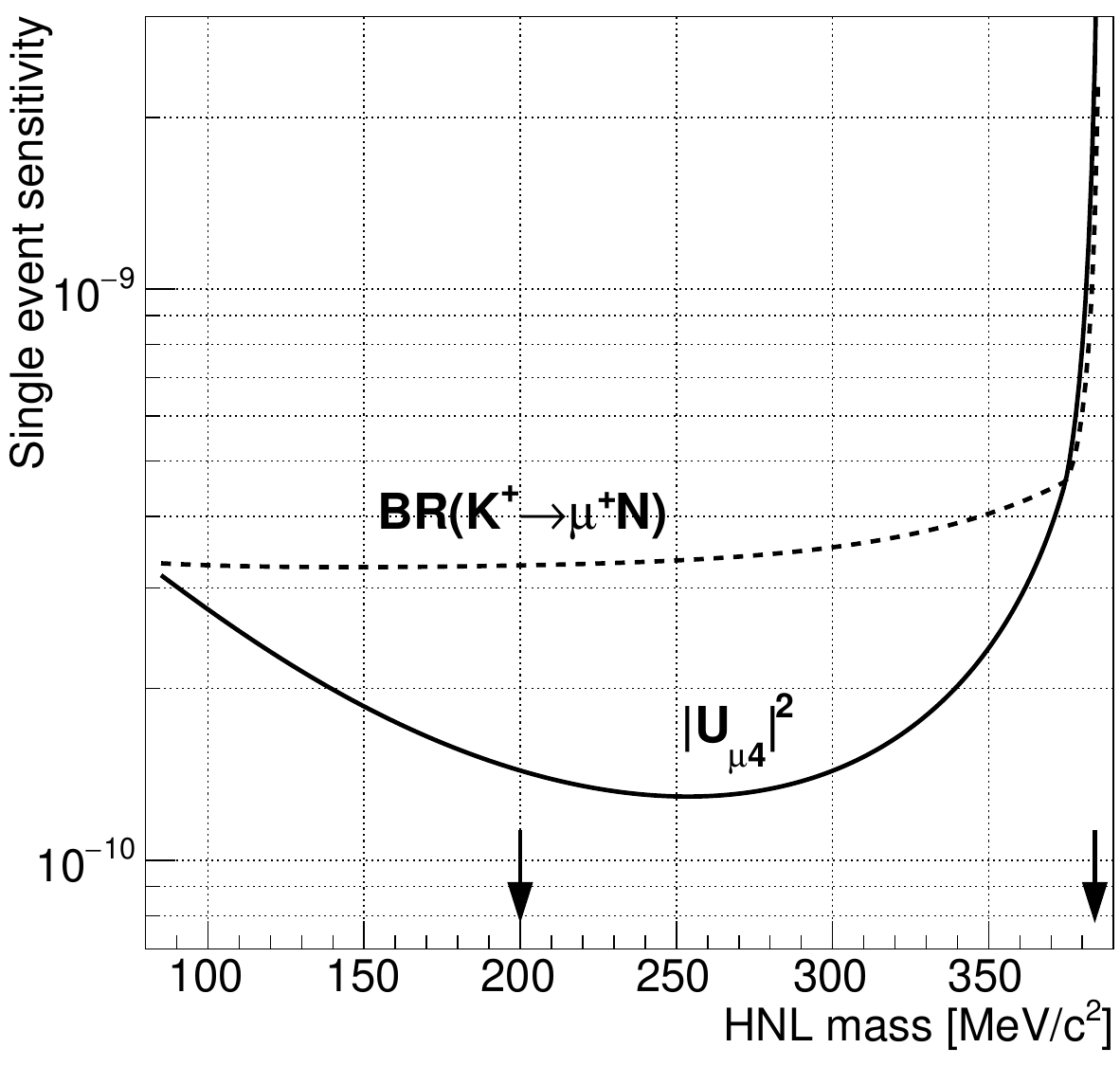}}%
\boldmath
\unboldmath
\end{center}
\vspace{-14mm}
\caption{Left: observed number of events $N_{\rm obs}$, observed upper limit at 90\% CL of the number of signal events $N_S$, and expected $\pm1\sigma$ and $\pm2\sigma$ bands of the upper limit in the null hypothesis for each HNL mass value considered. Right: single event sensitivity values of ${\cal B}_{\rm SES}(K^+\to\mu^+N)$ (dashed line) and $|U_{\mu4}|^2_{\rm SES}$ (solid line) as functions of the assumed HNL mass. Boundaries of the HNL search region are indicated by vertical arrows.}
\label{fig:ses}
\end{figure}

The number of observed events, $N_{\rm obs}$, within the signal window and the quantities $N_{\rm exp}$ and $\delta N_{\rm exp}$
are used to compute the local signal significance for each mass hypothesis.
It is found that the significance never exceeds 3 standard deviations, therefore no HNL production signal is observed.
Upper limits at 90\% CL of the number of $K^+\to\mu^+N$ decays, $N_S$, in each HNL mass hypothesis are evaluated from the quantities $N_{\rm obs}$, $N_{\rm exp}$ and $\delta N_{\rm exp}$ using the ${\rm CL_S}$ method~\cite{re02}. The values of $N_{\rm obs}$, the observed upper limits of $N_S$, and the expected $\pm1\sigma$ and $\pm 2\sigma$ bands of variation of $N_S$ in the null (i.e. background-only) hypothesis are shown in Fig.~\ref{fig:ses}~(left).

The single-event sensitivity (SES) branching fraction ${\cal B}_{\rm SES}(K^+\to\mu^+N)$ and mixing parameter values $|U_{\mu4}|^2_{\rm SES}$, corresponding to the observation of one signal event, are defined in each HNL hypothesis as
\begin{displaymath}
{\cal B}_{\rm SES}(K^+\to\mu^+N) = \frac{1}{N_K \cdot A_N} ~~~~ {\rm and} ~~~~
|U_{\mu 4}|^2_{\rm SES} = \frac{{\cal B}_{\rm SES}(K^+\to\mu^+N)}{{\cal B}(K^+\to\mu^+\nu) \cdot \rho_\mu(m_N)},
\end{displaymath}
with the kinematic factor $\rho_\mu(m_N)$ given in Eq.~(\ref{eq:rho}). They are shown as functions of the HNL mass in Fig.~\ref{fig:ses}~(right). The expected number of $K^+\to\mu^+N$ signal events, $N_S$, is written as
\begin{displaymath}
N_S = {\cal B}(K^+\to\mu^+N) / {\cal B}_{\rm SES}(K^+\to\mu^+N) = |U_{\mu 4}|^2/|U_{\mu 4}|^2_{\rm SES},
\end{displaymath}
which is used to obtain upper limits at 90\% CL of the branching fraction ${\cal B}(K^+\to\mu^+N)$ and the mixing parameter $|U_{\mu4}|^2$ from those of $N_S$.

\begin{figure}[p]
\begin{center}
\vspace{-3mm}
\resizebox{0.63\textwidth}{!}{\includegraphics{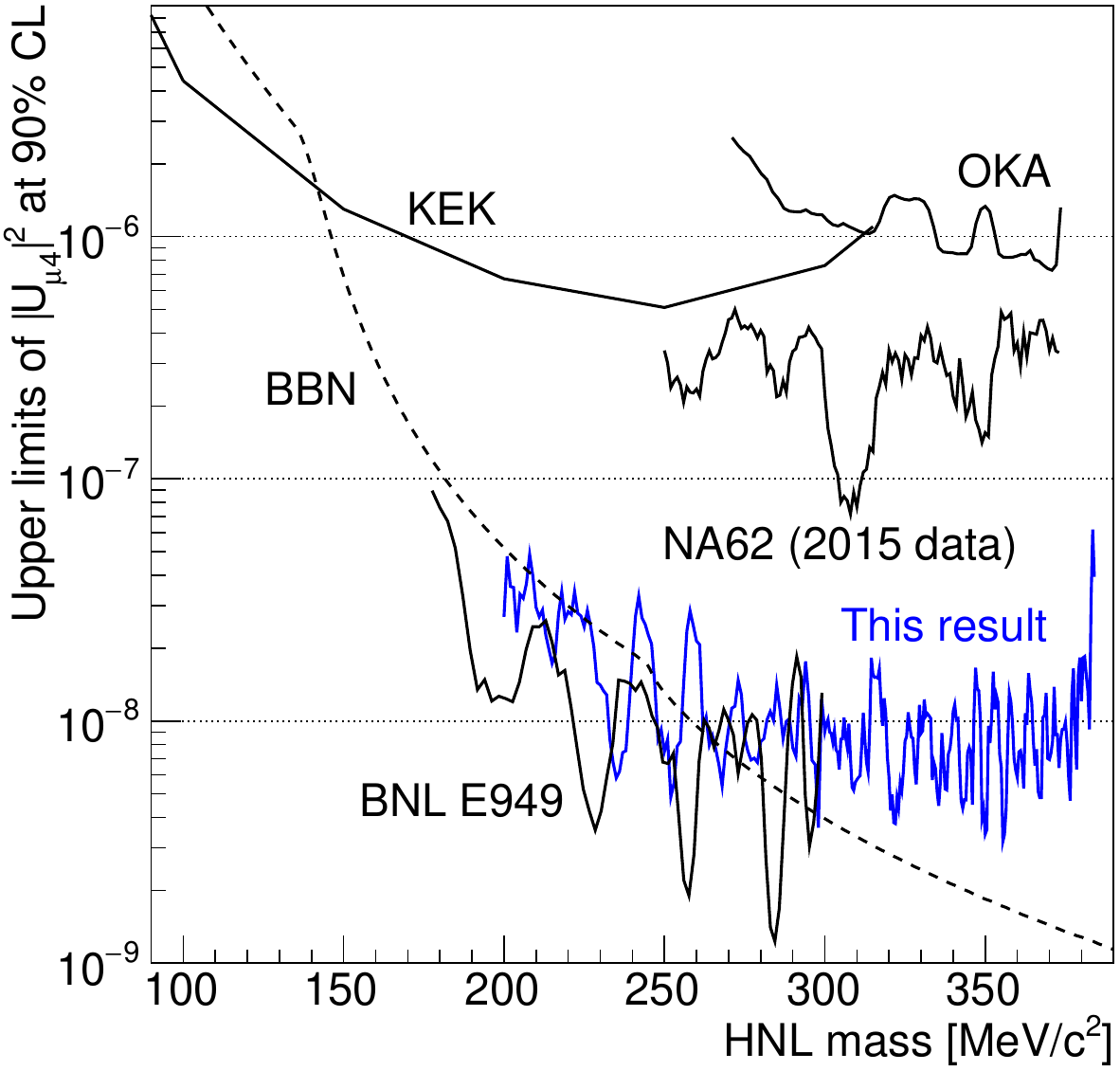}}
\end{center}
\vspace{-15mm}
\caption{Upper limits at 90\% CL of $|U_{\mu4}|^2$ obtained for each assumed HNL mass, compared to the upper limits established by earlier HNL production searches in $K^+\to\mu^+N$ decays at NA62~\cite{co18}, BNL-E949~\cite{ar15}, OKA~\cite{oka18} and KEK~\cite{ya84}. The lower boundary of $|U_{\mu4}|^2$ imposed by the BBN constraint~\cite{do00} is shown by a dashed line.}
\label{fig:world-mu}
\end{figure}

The upper limits obtained for $|U_{\mu4}|^2$ are compared with the results from earlier searches for the $K^+\to\mu^+N$ decay~\cite{co18,ar15,oka18,ya84}, and the Big Bang nucleosynthesis (BBN) constraint~\cite{do00}, in Fig.~\ref{fig:world-mu}. The results of the current study represent the first HNL production search in the mass range 374--384~MeV/$c^2$, and improve on previous NA62 results in the mass range 300--374~MeV/$c^2$~\cite{co18} by more than an order of magnitude. In the range 200--300~MeV/$c^2$, the sensitivity achieved is similar to that of the BNL-E949 experiment~\cite{ar15}.

\begin{figure}[p]
\begin{center}
\resizebox{0.63\textwidth}{!}{\includegraphics{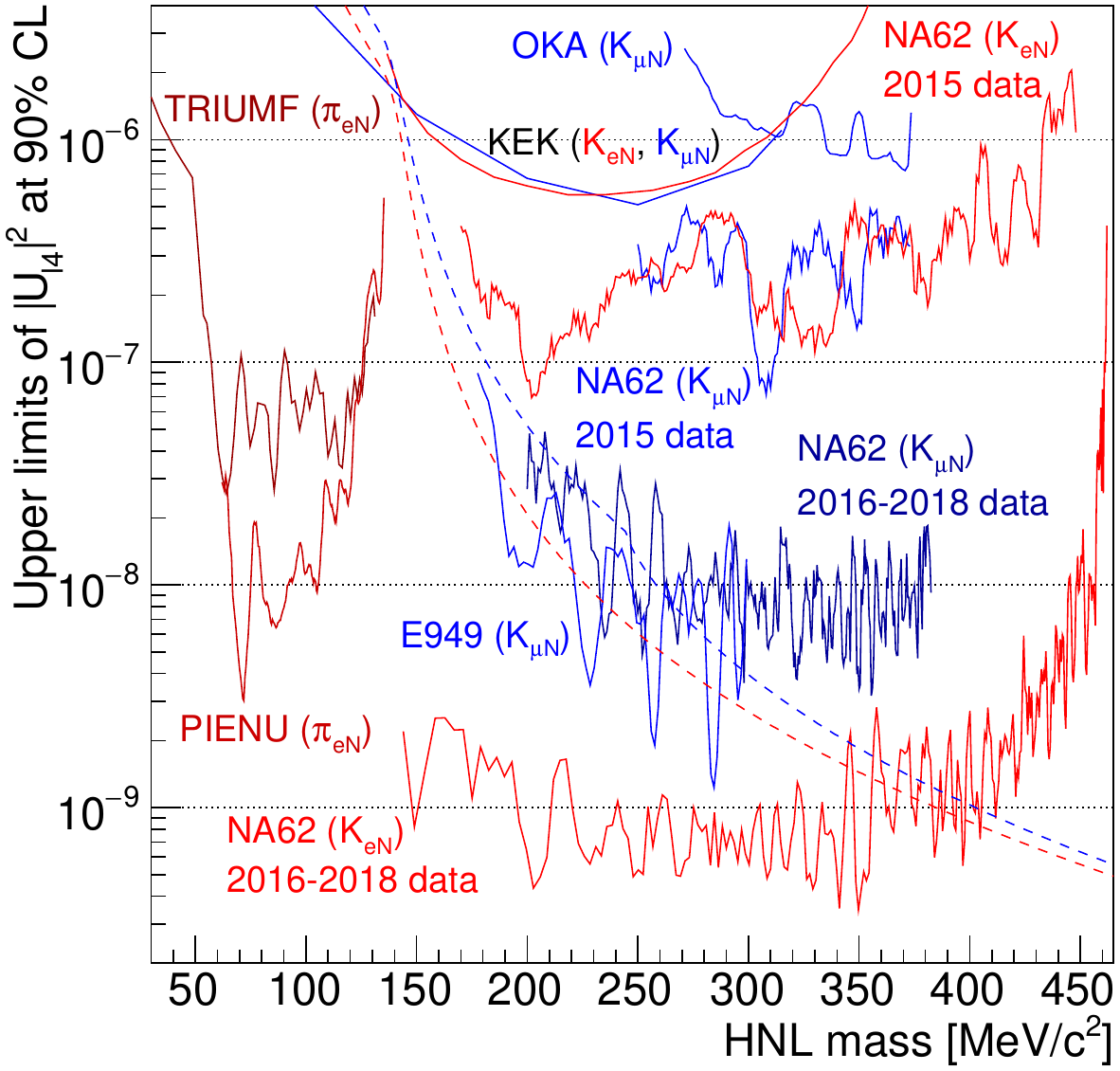}}
\end{center}
\vspace{-15mm}
\caption{Summary of upper limits at 90\% CL of $|U_{e4}|^2$ (red solid lines) and $|U_{\mu4}|^2$ (blue solid lined) obtained from HNL production searches in $K^+$ decays: this analysis, NA62~\cite{co18,hnl-e}, BNL-E949~\cite{ar15}, OKA~\cite{oka18}, KEK~\cite{ya84}; and in $\pi^+$ decays: TRIUMF~\cite{br92}, PIENU~\cite{ag17}. The lower boundaries of $|U_{e4}|^2$ and $|U_{\mu4}|^2$ imposed by the BBN constraint~\cite{do00} are shown by the lower and upper dashed lines, respectively.}
\label{fig:world}
\end{figure}

A comparison of the above upper limits of $|U_{\mu4}|^2$ with the upper limits of $|U_{e4}|^2$ obtained from HNL production searches in $K^+\to e^+N$~\cite{co18,hnl-e,ya84} and $\pi^+\to e^+N$~\cite{br92,ag17} decays is shown in Fig.~\ref{fig:world}. Upper limits of ${\cal O}(10^{-5})$ obtained on $|U_{\mu4}|^2$ in the mass range 16--34~MeV/$c^2$ from searches of the $\pi^+\to\mu^+N$ process~\cite{ag19} are not shown. In comparison to the limits of $|U_{\mu4}|^2$ obtained from direct HNL decay searches~\cite{ps191,abe19}, the limits from production searches are weaker but more robust because they are based on fewer theoretical assumptions.

\section{\boldmath Search for $K^+\to\mu^+\nu X$ and $K^+\to\mu^+\nu\nu\bar\nu$ decays}
\label{sec:kmx}

The $K^+\to\mu^+\nu X$ process is investigated in the framework of the scalar and vector mediator models, defined for non-zero mediator mass $m_X$~\cite{kr20}.
In total, 37 mass hypotheses equally spaced in the range 10--370~MeV/$c^2$ are examined. The $K^+\to\mu^+\nu\nu\bar\nu$ decay is investigated assuming the SM differential decay rate distribution~\cite{go16}.

The true missing mass spectrum lies in the $m_X\le m_{\rm miss}\le m_K-m_\mu$ range for the $K^+\to\mu^+\nu X$ decay, and in the $0\le m_{\rm miss}\le m_K-m_\mu$ range for the $K^+\to\mu^+\nu\nu\bar\nu$ decay (neglecting the neutrino mass). In both cases, a signal would manifest itself as an excess of data events over the estimated background at large reconstructed $m_{\rm miss}^2$ values as shown in Fig.~\ref{fig:mmiss2}~ (top-right). Therefore the event selection requires that $m_{\rm miss}^2>m_0^2$. The $m_0$ value is optimized to obtain the strongest expected upper limit of the decay rate in the null hypothesis, considering that signal acceptances and backgrounds both decrease as functions of $m_0^2$. The optimization is performed independently for each of the possible signals listed above.

\begin{figure}[t]
\begin{center}
\resizebox{0.5\textwidth}{!}{\includegraphics{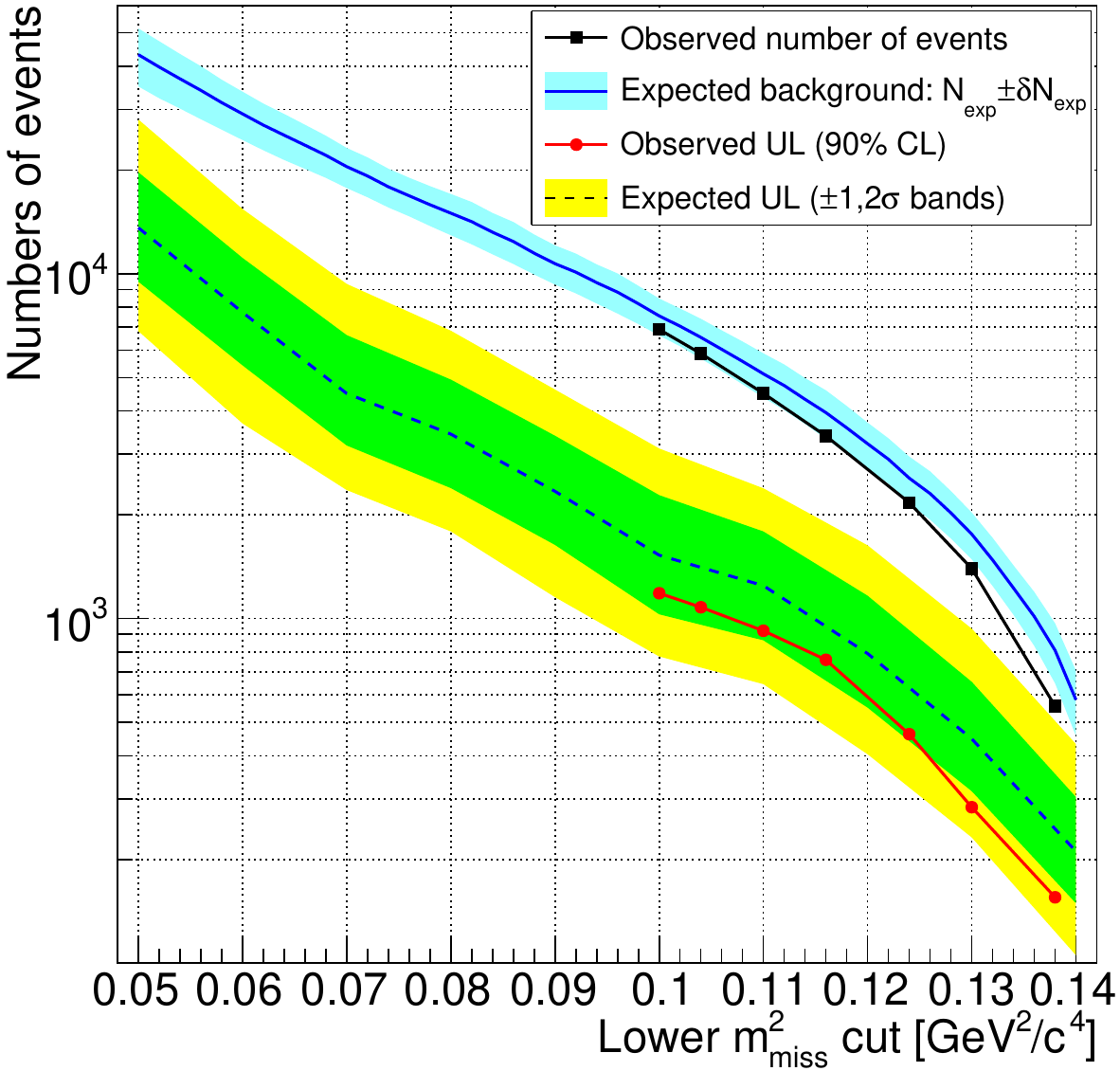}}%
\resizebox{0.5\textwidth}{!}{\includegraphics{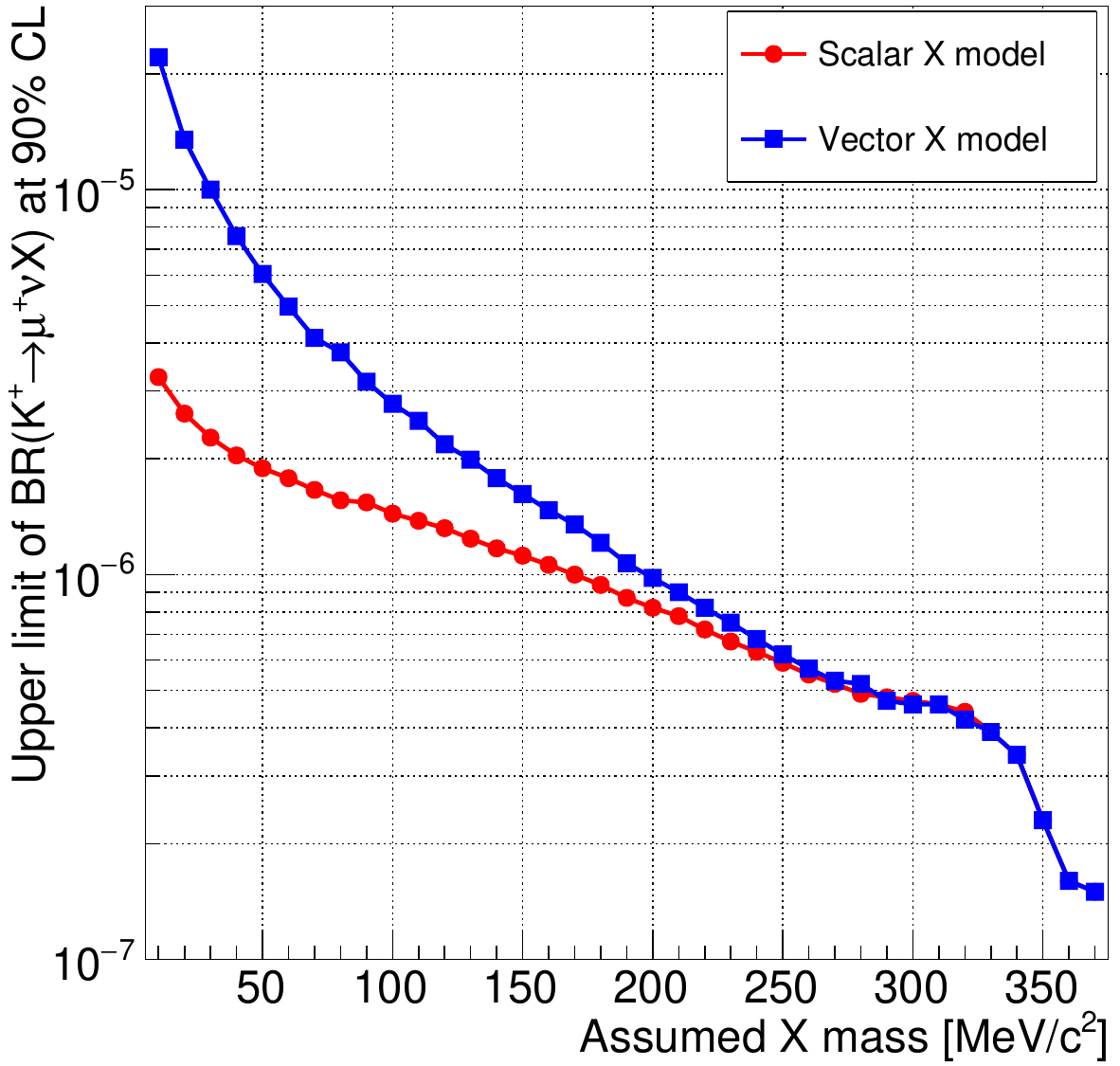}}
\end{center}
\vspace{-12mm}
\caption{Left: expected background, its uncertainty, and expected $\pm1\sigma$ and $\pm2\sigma$ bands of the upper limit on the number at 90\% CL of signal events $N_S$ in the null hypothesis, for each lower squared missing mass cut ($m_0^2$) considered to optimize the definition of the $K^+\to\mu^+\nu X$ and $K^+\to\mu^+\nu\nu\bar\nu$ signal regions. Observed numbers of events and upper limits of $N_S$ are shown for $m_0^2$ values found to be optimal for certain $m_X$ hypotheses. Right: upper limits of ${\cal B}(K^+\to\mu^+\nu X)$ obtained at 90\% CL for each $m_X$ hypothesis for the scalar and vector mediator models.}
\label{fig:kmunux}
\end{figure}

The numbers of background events, $N_{\rm exp}$, and their uncertainties, $\delta N_{\rm exp}$, estimated with simulations (Section~\ref{sec:bkg}) are shown as functions of $m_0^2$ in Fig.~\ref{fig:kmunux}~(left). Also shown are the expected upper limits at 90\% CL of the number of signal events, $N_S$, and their $\pm1\sigma$ and $\pm 2\sigma$ bands of variation in the null hypothesis, obtained from $N_{\rm exp}$ and $\delta N_{\rm exp}$ using the ${\rm CL_S}$ method~\cite{re02} for each $m_0^2$ value considered.


For the $K^+\to\mu^+\nu X$ decay in $m_X$ hypotheses of 320--370~{\rm MeV}/$c^2$, the signal region is defined $m_0^2=m_X^2$ (rounded up to the nearest multiple of $0.02~{\rm GeV}^2/c^4$), avoiding a significant loss of signal acceptance. For the $K^+\to\mu^+\nu X$ decay in $m_X$ hypotheses of 10--310~{\rm MeV}/$c^2$, and for the  $K^+\to\mu^+\nu\nu\bar\nu$ decay, the signal region is defined as $m_0^2=0.1~{\rm GeV}^2/c^4$. The background composition for this $m_0^2$ value is reported in Table~\ref{tab:bkg_km2}. Optimal sensitivity is obtained in this case with a reduced signal acceptance. In particular, the acceptance for the $K^+\to\mu^+\nu\nu\bar\nu$ decay decreases from $A_{\mu\nu\nu\nu}^0=0.277$ to $A_{\mu\nu\nu\nu}=0.103$.

The observed numbers of events and upper limits of $N_S$ for the above set of $m_0^2$ values are displayed in Fig.~\ref{fig:kmunux}~(left). Upper limits of ${\cal B}(K^+\to\mu^+\nu X)$ in the scalar and vector $X$ models as functions of the assumed $m_X$, obtained from those of $N_S$ similarly to the HNL case, are shown in Fig.~\ref{fig:kmunux}~(right). The limits obtained in the scalar model are stronger than those in the vector model due to the larger mean $m_{\rm miss}$ value.

In the search for the $K^+\to\mu^+\nu\nu\bar\nu$ decay, $N_{\rm obs}=6894$ events are observed in the signal region $m_{\rm miss}^2>0.1~{\rm GeV}^2/c^4$, with an expected background of $N_{\rm exp}=7549\pm928$ events. This leads to an observed (expected) upper limit at 90\% CL of 1184~(1526) events for the number of signal events $N_S$. An upper limit is established on the decay rate using the relation $N_S = N_K\cdot {\cal B}(K^+\to\mu^+\nu\nu\bar\nu)\cdot A_{\mu\nu\nu\nu}$:
\begin{displaymath}
{\cal B} (K^+\to\mu^+\nu\nu\bar\nu)<1.0\times 10^{-6} ~~~ {\rm at~90\%~CL},
\end{displaymath}
improving by a factor of 2.4 on the most stringent previous limit obtained by the BNL-E949 experiment~\cite{ar16}. Both this and BNL-E949 $K^+\to\mu^+\nu\nu\bar\nu$ results are obtained assuming the SM differential rate. However the reconstructed missing mass intervals analysed are complementary: $m_{\rm miss}>316~{\rm MeV}/c^2$ in this study, and $230<m_{\rm miss}<300~{\rm MeV}/c^2$ at BNL-E949.


\newpage

\section*{Summary}

A search for HNL production in $K^+\to\mu^+N$ decays has been performed using the data set collected by the NA62 experiment in 2016--2018. Upper limits of the HNL mixing parameter $|U_{\mu4}|^2$ are established at the level of ${\cal O}(10^{-8})$ over the HNL mass range of 200--384~MeV/$c^2$ with the assumption of mean lifetime exceeding 50~ns, improving on the previous HNL production searches. The first search for $K^+\to\mu^+\nu X$ decays has been performed, where $X$ is a scalar or vector hidden sector mediator in the mass range 10--370~MeV/$c^2$, which decays to an invisible final state. Upper limits obtained at 90\% CL on the decay branching fraction range from ${\cal O}(10^{-5})$ for low $m_X$ values to ${\cal O}(10^{-7})$ for high $m_X$ values. An upper limit of $1.0\times 10^{-6}$ is obtained at 90\% CL on the branching fraction of the $K^+\to\mu^+\nu\nu\bar\nu$ decay, assuming the SM differential decay rate, which improves on the earlier searches for this process.


\section*{Acknowledgements}

It is a pleasure to express our appreciation to the staff of the CERN laboratory and the technical staff of the participating laboratories and universities for their efforts in the operation of the experiment and data processing. We are grateful to Diego Redigolo and Kohsaku Tobioka for fruitful discussions and for the inputs provided on the $K^+\to\mu^+\nu X$ decay phenomenology.

The cost of the experiment and its auxiliary systems was supported by the funding agencies of
the Collaboration Institutes. We are particularly indebted to:
F.R.S.-FNRS (Fonds de la Recherche Scientifique - FNRS), Belgium;
BMES (Ministry of Education, Youth and Science), Bulgaria;
NSERC (Natural Sciences and Engineering Research Council), funding SAPPJ-2018-0017 Canada;
NRC (National Research Council) contribution to TRIUMF, Canada;
MEYS (Ministry of Education, Youth and Sports),  Czech Republic;
BMBF (Bundesministerium f\"{u}r Bildung und Forschung) contracts 05H12UM5, 05H15UMCNA and 05H18UMCNA, Germany;
INFN  (Istituto Nazionale di Fisica Nucleare),  Italy;
MIUR (Ministero dell'Istruzione, dell'Univer\-sit\`a e della Ricerca),  Italy;
CONACyT  (Consejo Nacional de Ciencia y Tecnolog\'{i}a),  Mexico;
IFA (Institute of Atomic Physics) Romanian CERN-RO No.1/16.03.2016 and Nucleus Programme PN 19 06 01 04,  Romania;
INR-RAS (Institute for Nuclear Research of the Russian Academy of Sciences), Moscow, Russia;
JINR (Joint Institute for Nuclear Research), Dubna, Russia;
NRC (National Research Center)  ``Kurchatov Institute'' and MESRF (Ministry of Education and Science of the Russian Federation), Russia;
MESRS  (Ministry of Education, Science, Research and Sport), Slovakia;
CERN (European Organization for Nuclear Research), Switzerland;
STFC (Science and Technology Facilities Council), United Kingdom;
NSF (National Science Foundation) Award Numbers 1506088 and 1806430,  U.S.A.;
ERC (European Research Council)  ``UniversaLepto'' advanced grant 268062, ``KaonLepton'' starting grant 336581, Europe.

Individuals have received support from:
Charles University Research Center (UNCE/SCI/ 013), Czech Republic;
Ministry of Education, Universities and Research (MIUR  ``Futuro in ricerca 2012''  grant RBFR12JF2Z, Project GAP), Italy;
Russian Foundation for Basic Research  (RFBR grants 18-32-00072, 18-32-00245), Russia;
Russian Science Foundation (RSF 19-72-10096), Russia;
the Royal Society  (grants UF100308, UF0758946), United Kingdom;
STFC (Rutherford fellowships ST/J00412X/1, ST/M005798/1), United Kingdom;
ERC (grants 268062,  336581 and  starting grant 802836 ``AxScale'');
EU Horizon 2020 (Marie Sk\l{}odowska-Curie grants 701386, 842407, 893101).



\end{document}